\documentclass[12pt]{article}
\pdfoutput=1
\usepackage{amsmath}
\usepackage{amssymb} 
\usepackage{bbm} 
\usepackage[small]{caption2} 
\usepackage{fleqn} 
\usepackage{graphicx} 
\usepackage{mathrsfs}   
\usepackage[small,loose]{subfigure}  
\usepackage{cite} 
\usepackage{color}

\newcommand{\eVdist}{\kern-0.06em}

\newcommand{\mev}{\:\text{Me\eVdist V}}
\newcommand{\gev}{\:\text{Ge\eVdist V}}

\newcommand{\be}{\begin{equation}}
\newcommand{\ee}{\end{equation}}
\newcommand{\bea}{\begin{align}}
\newcommand{\eea}{\end{align}}

\allowdisplaybreaks[1]

\title{Direct detection of dark matter in models with a light $Z'$}

\begin{document}

\begin{titlepage}

\vspace*{-3.0cm}
\begin{flushright}
OUTP-11-44
\end{flushright}


\begin{center}
{\large\bf
Direct detection of dark matter in models with a light $Z'$
}

\vspace{1cm}

\textbf{
Mads T.~Frandsen,
Felix Kahlhoefer,
Subir Sarkar,
Kai Schmidt-Hoberg
}
\\[5mm]
\textit{\small
Rudolf Peierls Centre for Theoretical Physics, University of Oxford,\\
1 Keble Road, Oxford OX1 3NP, UK
}
\end{center}

\vspace{1cm}

\begin{abstract}
  We discuss the direct detection signatures of dark matter
  interacting with nuclei via a new neutral gauge boson $Z'$,
  focussing on the case where both the dark matter and the $Z'$ have
  mass of a few GeV. Isospin violation (i.e.\ different couplings to
  protons and neutrons) arises naturally in this scenario. In
  particular it is possible to reconcile the preferred parameter
  regions inferred from the observed DAMA and CoGeNT modulations with
  the bounds from XENON100, which requires $f_n/f_p \simeq
  -0.7$. Moreover, the $Z'$ mediator can also yield a large
  spin-dependent cross-section which could 
  contribute to the DAMA
  signal, while the spin-independent cross-section is adequate to
  explain the CoGeNT signal.
\end{abstract}

\end{titlepage}

\tableofcontents
\newpage

\section{Introduction}

For many years, direct detection experiments have attempted to confirm
if dark matter (DM) is indeed made of new non-baryonic relic particles
\cite{Primack:1988zm,Gaitskell:2004gd}. They have ruled out various DM
candidates and severely constrained others, but have not yet been able
to unambiguously observe a DM signal. Nevertheless, recent data from
the DAMA \cite{Bernabei:2010mq} and CoGeNT
\cite{Aalseth:2010vx,Aalseth:2011wp} experiments have been interpreted
as being due to spin-independent (SI) scattering of DM particles with
relatively high cross-section ($\sim 10^{-40}$ cm$^2$) and small mass
($\sim 10$ GeV) \cite{Kelso:2010sj}. However, this explanation is in
strong tension with other null results, most notably from the CDMS
\cite{Ahmed:2010wy}, XENON10 \cite{Angle:2011th} and XENON100
\cite{Aprile:2011hi} experiments.

The analysis of direct detection experiments usually assumes that the
DM particle couples with equal strength to protons and neutrons.
However, this need not be the case, \emph{e.g.} if the mediator of the
DM scattering couples to isospin
\cite{Giuliani:2005my}. Isospin-dependent couplings occur naturally
for vector mediators, for example the photon couples only to protons,
while the $Z$ boson couples predominantly to neutrons. There can
even be cancellations between the scattering on protons and neutrons,
reducing the sensitivity of some direct detection experiments.

It has been noted that the tension between the DAMA/CoGeNT and XENON
results can be alleviated by considering such isospin-dependent
couplings, in particular a ratio of neutron to proton couplings
$f_n/f_p \simeq -0.7$
\cite{Kurylov:2003ra,Giuliani:2005my,Chang:2010yk,Feng:2011vu,Frandsen:2011ts,DelNobile:2011je,Schwetz:2011xm,Farina:2011pw,Fox:2011px,McCabe:2011sr}.
We demonstrate that this value can be obtained from a single new
vector mediator. Specifically we consider the case of a light $Z'$
with a mass that is comparable to that of the DM particle, \emph{i.e.}
several GeV. It mediates the interaction between the SM and a new
hidden sector which includes the DM particle and is uncharged under
the Standard $SU(2)_\mathrm{L} \times U(1)_Y$ Model (SM) but charged
under the new $U(1)$. In this framework it is possible to get both the
required value of $f_n/f_p$, as well as sufficiently high
cross-sections to account for the absolute signal levels observed by
DAMA and CoGeNT. Our analysis can also be applied to the composite
vector particles arising in new strong dynamics models of DM.

The outline of this paper is as follows: first we discuss how to
obtain the effective coupling constants $f_n$ and $f_p$ in the general
case of a vector mediator which mixes with the neutral gauge bosons of
the SM. We then study, using an effective Lagrangian, how the required
value of $f_n/f_p$ can be realised with such a $Z'$. We discuss other
experimental bounds on our model parameters and show that the required
cross-section needed to explain the DAMA and CoGeNT signals, with
$f_n/f_p\simeq -0.7$, remains viable. Finally we note that data from
collider experiments such as BaBar, Belle, BEPC and LHCb will be able
to detect or rule out such a light $Z'$, as discussed in \emph{e.g.}
Ref.~\cite{Reece:2009un}.

\section{DM interaction via a vector mediator $R$}
\label{sec:DMinteraction}

In this section we discuss the general calculation of the ratio of the
effective coupling constants for DM scattering on neutrons and protons
$f_n/f_p$ via a vector mediator. For this purpose, we first consider
an (axial)-vector mass-eigenstate $R$ interacting with the SM fermions
$f$ and the fermionic DM $\chi$ via the neutral current Lagrangian
\begin{align}
  {\cal L}_{R}^\mathrm{NC} = & R_{\mu} \bar{\chi} \gamma^{\mu}(
  g^\mathrm{V}_\chi - g^\mathrm{A}_\chi \gamma^5) \chi + R_\mu
  \bar{f} \gamma^\mu (g^\mathrm{V}_{f} - g^\mathrm{A}_{f} \gamma^5) f
  \; .
\end{align}
Integrating out $R$ generates the effective four-fermion interactions
\begin{align}
  {\cal L}_{R}^\mathrm{eff} = & \, b^\mathrm{V}_{f} \bar{\chi} \gamma_\mu
  \chi \bar{f} \gamma^\mu f\, + \, b^\mathrm{A}_{f} \bar{\chi} \gamma_\mu
  \gamma^5 \chi \bar{f} \gamma^\mu
  \gamma^5 f  \nonumber \\
  & + c^1_{f} \bar{\chi} \gamma_\mu \gamma^5 \chi \bar{f} \gamma^\mu
  f\, + \, c^2_{f} \bar{\chi} \gamma_\mu \chi \bar{f} \gamma^\mu
  \gamma^5 f\, \; ,
\label{eq:bf}
\end{align}
where $b^\mathrm{V,A}_{f} = g^\mathrm{V,A}_{\chi}
g^\mathrm{V,A}_{f}/m_R^2$ and $c^{1,2}_{f} = g^\mathrm{A,V}_{\chi}
g^\mathrm{V,A}_{f}/m_R^2$.  In the non-relativistic limit
relevant for the interaction between DM
and detector nuclei, the mixing terms proportional to $c^i_{f}$ are
suppressed, so we neglect them in our analysis.

We assume that $R$ arises from the mixing of an interaction eigenstate
vector $X$ with the SM $U(1)_Y$ $B$ field and the neutral component
$W^3$ of $SU(2)_\mathrm{L}$ weak fields
\begin{align}
\label{eq:general-mixing}
\left(\begin{array}{c} B_\mu \\ W^3_\mu \\ X_\mu \end{array}\right)=
\left(\begin{array}{ccc}  N_{11} & N_{12} & N_{13}
\\ N_{21} & N_{22} & N_{23}
\\ N_{31} & N_{32} & N_{33}
\end{array}\right)
\left(\begin{array}{c} A_\mu \\ Z_\mu \\ R_\mu \end{array}\right).
\end{align}
Here $A$, $Z$ are the physical photon and neutral massive
gauge boson fields of the SM. As we discuss below, there can also be a
kinetic mixing between $X$ and the SM gauge bosons in addition to the
mass mixing described here.  However, by redefining the fields to have
standard kinetic terms, this kinetic mixing is \emph{equivalent} to an
additional mass mixing contribution, \emph{cf.}
Sec.~\ref{sec:zprime}.

In addition to any direct couplings of $X$ to SM fermions, denoted by
$f^\mathrm{V, A}$, the mixing will introduce additional terms, so that
the overall couplings of $R$ to light quarks will be given in terms of
the mass mixing matrix, as
\begin{align}
  g_{u}^\mathrm{V} &= -\frac{1}{12}(5 {\hat g'} N_{13}+3 {\hat
    g}N_{23})-f_u^\mathrm{V} N_{33} \ , & g_{u}^\mathrm{A} &=
  \frac{1}{4}({\hat g'} N_{13} - {\hat g}N_{23})-f_u^\mathrm{A} N_{33}
  \ ,
  \nonumber \\
  g_{d}^\mathrm{V} &= \frac{1}{12}({\hat g'} N_{13}+3 {\hat
    g}N_{23})-f_d^\mathrm{V} N_{33} \ , & g_{d}^\mathrm{A} &= -\frac{1}{4}
  ({\hat g'} N_{13} - {\hat g}N_{23})-f_d^\mathrm{A} N_{33}\; ,
\label{couplings}
\end{align}
where the numerical coefficients are determined from the hypercharge
and weak quantum numbers of the quarks. Similarly, the effective
vector and axial couplings of $R$ to the DM particle $\chi$ are given
by
\begin{align}
  g_{\chi}^\mathrm{V} &= f_\chi^\mathrm{V} N_{33} \ , & 
  g_{\chi}^\mathrm{A} &= f_\chi^\mathrm{A} N_{33} \;.
\label{DMcouplings}
\end{align}
In the case where the direct couplings between fermions and $X$ arise
from a charge under some new gauge group, the coupling constants
$f^\mathrm{V, A}$ will be given by the product of the corresponding
gauge coupling $g_X$ and the respective charge.

A mass mixing as in Eq.~\eqref{eq:general-mixing} not only induces
couplings of $R$ to the SM fermions, but also couplings of $Z$ (and
in some cases $A$) to the DM particle. The corresponding
coupling constants can be calculated analogously, by examining the
corresponding column of the mixing matrix. The effective coupling
constants for $A$ are obtained from the first column, \emph{i.e.}
replacing $N_{i3}$ by $N_{i1}$, and the effective coupling constants
for $Z$ are obtained from the second column. Note that if $N_{31} =
0$, there is no coupling of the physical photon to the DM state $\chi$
and so there are \emph{no DM millicharges}. In this case, the
couplings between $\chi$ and the SM fermions are given by
\begin{align}
  b^\mathrm{A,V}_{f} = b^\mathrm{A,V}_{f R} + b^\mathrm{A,V}_{f Z} =
  \frac{g^\mathrm{A,V}_{\chi R} g^\mathrm{A,V}_{f R}}{m_R^2} +
  \frac{g^\mathrm{A,V}_{\chi Z} g^\mathrm{A,V}_{f Z}}{m_Z^2} \; .
\end{align}

Let us for now focus on the induced effective vector-vector
interaction between the DM particle and nucleons $(p,n)$
\begin{align}
  {\cal L}^\mathrm{V} = \, f_{p} \bar{\chi} \gamma_\mu \chi
  \bar{p}\gamma^\mu p + f_{n} \bar{\chi} \gamma_\mu \chi \bar{n}
  \gamma^\mu n\, , \quad
\end{align}
with coefficients $f_{p,n}$ given by 
\begin{align}
  f_p = 2 b^\mathrm{V}_{u} + b^\mathrm{V}_{d} \ , \ f_n = 2
  b^\mathrm{V}_{d} + b^\mathrm{V}_{u} \, .
\end{align}
Note that because of the conservation of the vector current, there is
no contribution of sea quarks or gluons to the effective couplings.

\begin{figure}[!htp]
\begin{center}
{
\includegraphics[width=.45\columnwidth]{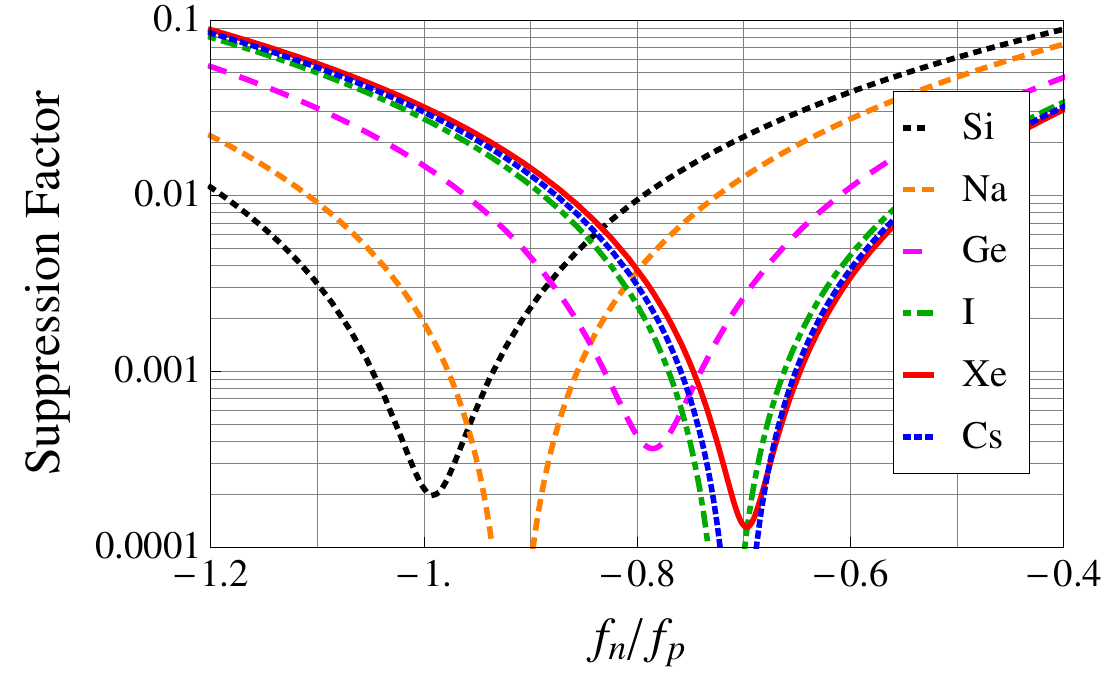}
\includegraphics[width=.45\columnwidth]{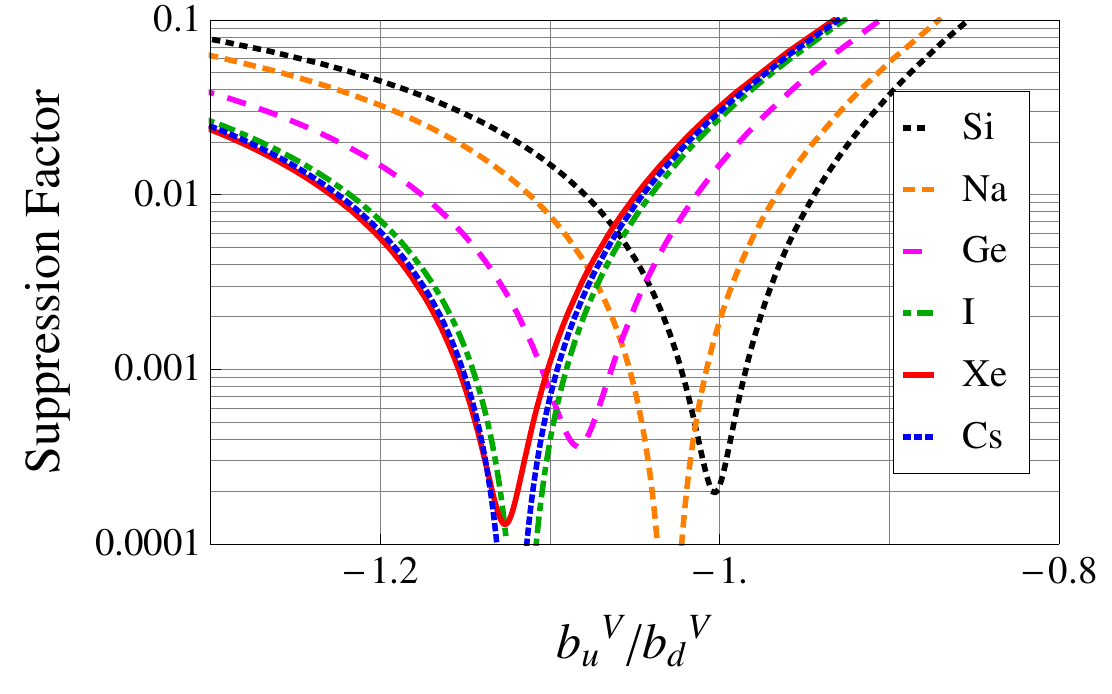}
}
\caption{The factor by which the cross-section for dark matter
  scattering on various isotopes is suppressed (compared to the
  standard case where $f_n = f_p$), as a function of the ratio
  $f_n/f_p$ (left) and the corresponding quark couplings $b_u^V/g_d^V$
  (right). In all cases, the natural isotopic abundance ratios have
  been assumed for the detector material.}
\label{fig1}
\end{center}
\end{figure}

Ultimately, we are interested in $\chi$ scattering off nuclei $N$ with
charge $\mathcal{Z}$ and mass number $\mathcal{A}$.  In the limit of zero momentum
transfer, the DM particle will interact coherently with the entire
nucleus $N$, resulting in the DM-nucleus cross-section
\begin{align}
  \sigma_N = \frac{\mu^2_{\chi N}}{\mu^2_{\chi n}}\left(\mathcal{Z} \frac{f_p}{f_n} 
  + \mathcal{A} - \mathcal{Z}\right)^2 \sigma_n \; , 
\end{align}
with $\sigma_n = \mu^2_{\chi n} f_n^2 /\pi$ the DM-neutron cross
section and $\mu_{XY}$ the reduced mass of the $(X,Y)$ system.  If the
DM scattering satisfies $f_n/f_p=\mathcal{Z}/(\mathcal{Z}-\mathcal{A})$ for a given nuclear isotope,
the corresponding cross-section is \emph{zero} and this isotope will
then not contribute to the constraint on DM-nucleon scattering.
Xenon, which typically gives the strongest constraint on the
DM-nucleon cross-section in the case of $f_n/f_p=1$ has $\mathcal{Z}=54$, while
$\mathcal{A}$ varies between 74 and 80. Consequently, if $-0.72 < f_n/f_p <
-0.68$ (corresponding to $-1.14 < b^\mathrm{V}_u/b^\mathrm{V}_d <
-1.11$), the scattering cross-section for DM particles on xenon nuclei
will be significantly suppressed. In Fig.~\ref{fig1}, we show the
suppression of the signal for different targets as a function of
$f_n/f_p$ and $b^\mathrm{V}_u/b^\mathrm{V}_d$.

At this point, it is instructive to look at a few familiar cases
\begin{enumerate}
\item For a mediator coupling to the baryonic current we
  have $b^\mathrm{V}_u/b^\mathrm{V}_d = f_n/f_p= 1$.
\item For a mediator coupling to the weak isospin current we have
  $b^\mathrm{V}_u/b^\mathrm{V}_d = f_n/f_p= -1$.
\item For a coupling to the EM current we have
  $b^\mathrm{V}_u/b^\mathrm{V}_d=-2$ and thus $f_n/f_p=0$.
\item Finally, a coupling to the vectorial part of the SM $Z$ current
  gives $b^\mathrm{V}_u/b^\mathrm{V}_d\sim -1/2$ and thus $f_n/f_p\gg
  1$, corresponding to a coupling dominantly to neutrons.
\end{enumerate}

In isolation, none of the possibilites above yields the ratio of
$f_n/f_p$ required to adequately suppress the DM scattering signal
from xenon. However as we will show in the following section, a single
light vector mediator which mixes with the neutral SM gauge bosons can
generate $f_n/f_p \simeq -0.7$ with sufficiently large cross-sections
to explain both the DAMA and CoGeNT signals, \emph{and} be in
agreement with all other experimental constraints. Alternatively there
may be several independent mediators which interfere to give the
desired value for $f_n/f_p$~\cite{Belyaev:2010kp,DelNobile:2011je}.

\section{Effective Lagrangian for a light $Z'$}
\label{sec:zprime}

From now on we will take the interaction eigenstate $X$ to be the new
gauge boson corresponding to an additional $U(1)_X$
symmetry~\cite{Holdom:1985ag,Babu:1997st,Cassel:2009pu,Hook:2010tw}. The
corresponding mass eigenstate $Z'$ is then a specific realisation of
the vector mediator $R$. The general possibility of DM (in particular
light DM) coupling to the SM via a $Z'$ has been considered
earlier~\cite{Mambrini:2010dq,Kang:2010mh,Chun:2010ve,Fox:2011qd,Gondolo:2011eq, Mambrini:2011dw,Mambrini:2011pw}.

We consider the following effective Lagrangian, which includes kinetic
mixing and mass mixing \cite{Babu:1997st}:\footnote{There could also
  be a mass mixing term between $Z'$ and $B$, which arises \emph{e.g.}\ in
  St\"uckelberg models. However, we consider here only such mass
  mixing as can be induced by a Higgs vev, which mixes only $Z$ and
  $Z'$.}
\begin{align}
  {\cal L} =& \; {\cal L}_{SM}
  -\frac{1}{4}\hat{X}^{\mu\nu}\hat{X}_{\mu\nu} + {1\over2} m_{\hat
    X}^2 \hat{X}_\mu \hat{X}^\mu - m_\chi \bar{\chi}\chi
  \nonumber  \\
  & - {1\over2} \sin \epsilon\, \hat{B}_{\mu\nu} \hat{X}^{\mu\nu} +\delta
  m^2 \hat{Z}_\mu \hat{X}^\mu - \sum_{f} f_f^\mathrm{V} \hat{X}^\mu
  \bar{f}\gamma_\mu f - f^\mathrm{V}_\chi \hat{X}^\mu
  \bar{\chi}\gamma_\mu \chi \;,
\label{LZprime}
\end{align}
where the $U(1)_X$ is assumed to be broken and the corresponding
vector boson mass is $m_{\hat X}$.  We denote fields in the original
basis with hats ($\hat B, \hat W^3, \hat X$) and define $\hat{Z}\equiv
\hat{c}_\mathrm{W} \hat{W}^3- \hat{s}_\mathrm{W} \hat{B}$, where
$\hat{s}_\mathrm{W} \, (\hat{c}_\mathrm{W})$ is the sine (cosine) of
the Weinberg angle and $\hat g', \, \hat g$ are the corresponding
gauge couplings.  Canonically normalised interaction eigenstates after
kinetic diagonalisation and normalisation are denoted without hats,
while the mass eigenstates after mass diagonalisation are denoted by
($A,Z,Z'$).  In the following we will abbreviate $\sin \theta \equiv
s_\theta, \cos \theta \equiv c_\theta, \tan\theta \equiv t_\theta$.

The diagonalisation of the above Lagrangian is discussed in detail in
\emph{e.g.} Ref.\cite{Babu:1997st}.  The field strengths are diagonalised
and canonically normalised via the following two consecutive
transformations
\begin{align}
\label{Zpmixing}
\left(\begin{array}{c} \hat B_\mu \\ \hat W_\mu^3 \\ \hat
    X_\mu \end{array}\right) & = \left(\begin{array}{ccc} 1 & 0 &
    -t_\epsilon \\ 0 & 1 & 0 \\ 0 & 0 &
    1/c_\epsilon \end{array}\right)
\left(\begin{array}{c} B_\mu \\ W_\mu^3 \\ X_\mu \end{array}\right) \ , \\
\left(\begin{array}{c} B_\mu \\ W_\mu^3 \\ X_\mu \end{array}\right) &
= \left(\begin{array}{ccc}
    \hat c_\mathrm{W} & -\hat s_\mathrm{W} c_\xi &  \hat s_\mathrm{W} s_\xi \\
    \hat s_\mathrm{W} & \hat c_\mathrm{W} c_\xi & - \hat c_\mathrm{W} s_\xi \\
    0 & s_\xi & c_\xi
\end{array} \right) 
\left(\begin{array}{c} A_\mu \\ Z_\mu \\ Z'_\mu \end{array}\right)
 \; ,
\end{align}
where
\begin{align}
  t_{2\xi}=\frac{-2c_\epsilon(\delta m^2+m_{\hat Z}^2 \hat
    s_\mathrm{W} s_\epsilon)} {m_{\hat X}^2-m_{\hat
      Z}^2 c_\epsilon^2 +m_{\hat Z}^2\hat s_\mathrm{W}^2 s_\epsilon^2
    +2\,\delta m^2\,\hat s_\mathrm{W} s_\epsilon} \; .
\label{eq:xi}
\end{align}
Multiplying the two matrices, we obtain
\begin{align}
  N_{13} =  \hat s_\mathrm{W} s_\xi - c_\xi t_\epsilon , \quad 
  N_{23} =  - \hat c_\mathrm{W} s_\xi \ , \quad
  N_{33} = c_\xi / c_\epsilon \; .
\end{align}
The resulting coupling structure of $Z'$ to the DM $\chi$ and SM
fermions can be written as
\begin{align}
  {\cal L}_{Z'} = & -{e\over 2 \hat c_\mathrm{W} {\hat s_\mathrm{W}}
  }\, c_\xi\, Z'_\mu\, \bar{f} \gamma^\mu \left\{ \left[ T_3^f ({\hat
        s_\mathrm{W}} t_\epsilon-t_\xi) + Q^f ({\hat s_\mathrm{W}}^2
      t_\xi - {\hat s_\mathrm{W}} t_\epsilon) \right]
  \right\} f  \nonumber \\
  & - \sum_{f} f^\mathrm{V}_f\frac{c_\xi}{c_\epsilon} Z'_\mu\, \bar{f}
  \gamma^\mu f \,-\, f_\chi^\mathrm{V} \, {c_\xi\over c_\epsilon} 
  Z'_\mu\,\bar{\chi}\gamma^\mu\chi \,,\label{Xfermion}
\end{align}
where the first term is due to mixing, with $T_3^f$ and $Q^f$ being
the weak isospin and electromagnetic charge of the fermion $f$, and
the second and third terms correspond to the rescaled direct
couplings. Thus, the effective couplings between $Z'$ and quarks are
\begin{align}
  g_{u}^\mathrm{V} &= \frac{5}{12}\frac{e c_\xi t_\epsilon}{\hat
    c_\mathrm{W}} +\frac{1}{4}\frac{e s_\xi t_\epsilon}{\hat
    c_\mathrm{W}\hat s_\mathrm{W}}-\frac{2}{3} \frac{e \hat
    s_\mathrm{W} s_\xi}{\hat c_\mathrm{W}}
  -\frac{c_\xi}{c_\epsilon}\,f_u^\mathrm{V} \ , 
  \nonumber \\
  g_{d}^\mathrm{V} &= -\frac{1}{12}\frac{e c_\xi t_\epsilon}{\hat
    c_\mathrm{W}} -\frac{1}{4}\frac{e s_\xi t_\epsilon}{\hat
    c_\mathrm{W}\hat s_\mathrm{W}}+\frac{2}{3} \frac{e \hat
    s_\mathrm{W} s_\xi}{\hat c_\mathrm{W}}
  -\frac{c_\xi}{c_\epsilon}\,f_d^\mathrm{V} \; .
\label{vectorcouplings}
\end{align}
Note also that in the case we consider, $N_{31} = 0$, which implies
that there is \emph{no DM-photon coupling}. This result is due to the
particular choice of mass mixing of $\hat X$ with only the $\hat Z$
interaction eigenstate.  If the mass mixing were to include a
component with the interaction eigenstate photon $\hat A$, then
millicharges for the DM would result as discussed \emph{e.g.} in
Ref.\cite{Feldman:2007wj}. In our case, an important consequence is
that even if we were to charge the SM fermions under $X$ as in
\emph{e.g.} Ref.\cite{Babu:1997st}, the electric charge remains
\emph{unchanged}, such that $\hat e =e$.  The physical $Z$ and $Z'$
masses after diagonalisation are given by
\begin{align} 
  m_Z^2 & = m_{\hat Z}^2 (1+{\hat s}_\mathrm{W} t_\xi
  t_\epsilon)+\delta m^2
  c_\epsilon^{-1} t_\xi \ , 
  \\
  m_{Z'}^2 & = \frac{m_{\hat X}^2 + \delta m^2 ({\hat s}_\mathrm{W}
    s_\epsilon-c_\epsilon t_\xi)}{c_\epsilon^2 (1+{\hat s}_\mathrm{W}
    t_\xi t_\epsilon)} \; .
\end{align}
For the ``physical'' weak angle, we adopt the definition
\begin{align}
  s_\mathrm{W}^2 c_\mathrm{W}^2=\frac{\pi\alpha(m_{Z})}{\sqrt{2}
    G_\mathrm{F} m_{Z}^2} \; .
\label{eq:swcw}
\end{align}

Eq.~\eqref{eq:swcw} is also true with the replacements
$s_\mathrm{W}\to\hat s_\mathrm{W}$, $c_\mathrm{W}\to\hat c_\mathrm{W}$
and $m_{Z}\to m_{\hat Z}$, leading to the identity
$s_\mathrm{W}c_\mathrm{W} m_{Z}=\hat s_\mathrm{W}\hat c_\mathrm{W}
m_{\hat Z}$.  From these equations we fix $\hat s_\mathrm{W}$ and
$m_{\hat Z}$ such that the experimentally well-measured quantities
$G_\mathrm{F}$ (or alternatively $s_\mathrm{W}$) and $m_{Z}$ come out
correctly.  From the mixing matrix $N_{ij}$, the nucleon-DM couplings
can be calculated using the formulae from Sec.~\ref{sec:DMinteraction}
\cite{Chun:2010ve}
\begin{align}
  f_p =& {\hat{g} f^\mathrm{V}_\chi \over 4 \hat c_{W} }
  {c_\xi^2\over c_\epsilon} t_\xi \left[ (1-4{\hat
      s_\mathrm{W}}^2)\left({1\over m_Z^2} - {1\over m_{Z'}^2}\right)
    -3 {\hat s_\mathrm{W}} {t_\epsilon\over t_\xi} \left({t_\xi^2\over
        m_Z^2} + {1\over m_{Z'}^2}\right) \right] -\frac{c_\xi^2}{c_\epsilon^2}
  f^\mathrm{V}_\chi \frac{2 f_u^\mathrm{V}+
    f_d^\mathrm{V}}{ m_{Z'}^2} \;,
  \nonumber\\
  f_n =& - {\hat{g} f^\mathrm{V}_\chi \over 4 \hat c_{W} }
  {c_\xi^2\over c_\epsilon} t_\xi \left[ \left({1 \over m_Z^2} -
      {1\over m_{Z'}^2}\right) + {\hat s_\mathrm{W}} {t_\epsilon \over
      t_\xi} \left({t_\xi^2 \over m_Z^2} + {1\over m_{Z'}^2}\right)
  \right] -\frac{c_\xi^2}{c_\epsilon^2} f^\mathrm{V}_\chi \frac{f_u^\mathrm{V}+2
    f_d^\mathrm{V}}{ m_{Z'}^2} \;.
\end{align}

Now that we have established the formalism, let us briefly come back
to the low energy Lagrangian from Eq.~\eqref{LZprime}.  There are
three terms which are of particular importance for the phenomenology
\begin{enumerate}
\item the direct  fermion couplings $\sum_{f} f_f^\mathrm{V} \hat{X}^\mu
  \bar{f}\gamma_\mu f$ ,
\item the kinetic mixing term ${1\over2} s_\epsilon\, \hat{B}_{\mu\nu}
  \hat{X}^{\mu\nu}$ , and
\item the mass mixing term $\delta m^2 \hat{Z}_\mu \hat{X}^\mu$ .
\end{enumerate}
In the following we will only consider tree-level couplings to quarks,
since couplings to leptons have to be strongly suppressed for the $Z'$
mass range we are interested in here.  In order to allow for the
standard Yukawa couplings, the Higgs field has to be uncharged as
well.  This leaves us with two possible scenarios for the $Z'$.  If
the $Z'$ couples to SM states, it has to couple to the baryon current
and hence corresponds to a gauged version of the $U(1)_B$ (a
\emph{baryonic $Z'$}).  The other possibility is that the SM is
completely uncharged under the $U(1)_X$ (a \emph{dark $Z'$}).  In both
cases there has to be an additional Higgs field $h'$ which gives mass
to the $Z'$. We discuss both cases below.
 
Coming to the kinetic and mass mixing terms, one might wonder how they
are generated in these setups.  It is well known that kinetic mixing
will be zero at tree level if both $U(1)$'s arise from the breaking of
a simple group \cite{Babu:1996vt}. However, if there is matter which
is charged under both $U(1)$'s, kinetic mixing will in general be
induced at 1-loop. If there are fields which are charged under both
$U(1)$'s, the mass mixing term can also be generated, \emph{e.g.} via the
operator $\tfrac{1}{\Lambda^2}h^\dagger D_\mu h h'^\dagger D^\mu h'
\rightarrow \tfrac{v^2 v'^2}{\Lambda^2}Z Z'$.  Note however that while the
kinetic mixing term is renormalisable, we have to invoke higher
dimensional operators to generate a mass mixing term, unless one of
the Higgs fields is charged under both the SM gauge group and the new
$U(1)_X$.

\subsection{Dark $Z'$}

\begin{figure}[!htp]
\begin{center}
{
\includegraphics[width=.45\columnwidth]{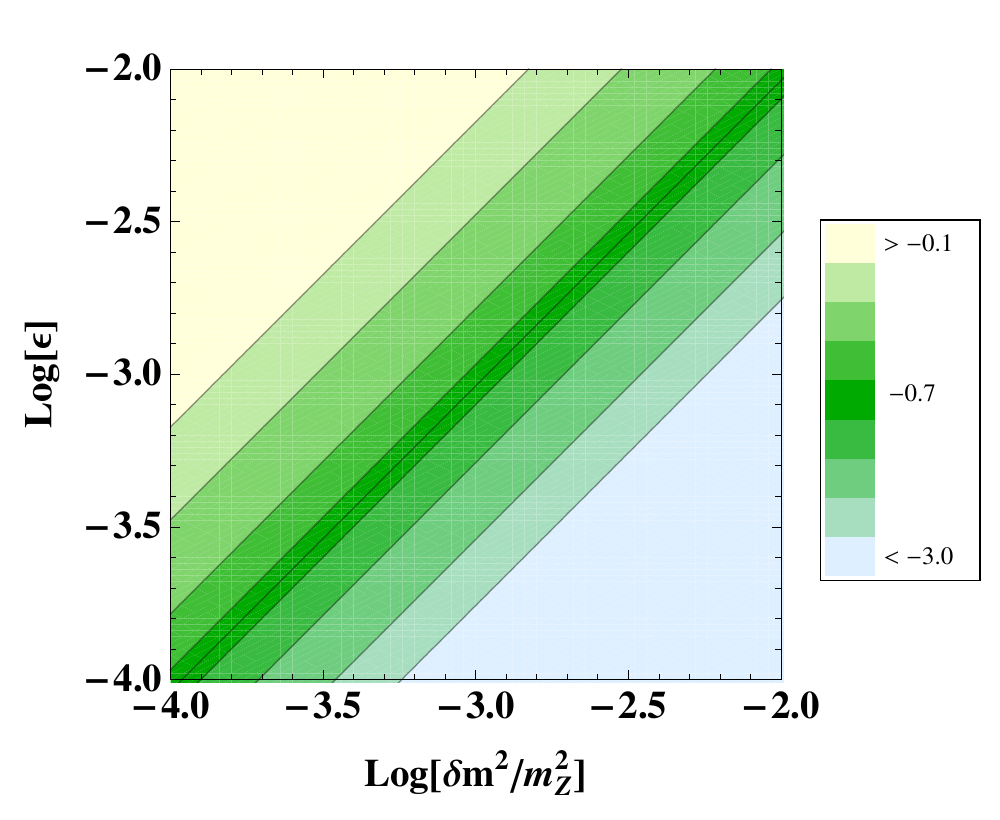}
}
\caption{Contours of $f_n/f_p$ in the plane of kinetic and mass mixing
  parameters $\epsilon$ and $\delta m$, with the dark green trough
  corresponding to $f_n/f_p \simeq -0.7$.}
\label{fig2}
\end{center}
\end{figure}

Let us first consider the case of a dark $Z'$, where the SM fields are
uncharged under the new gauge group.  If we also set $\delta m^2 = 0$,
\emph{i.e.}\ consider kinetic mixing alone, the mass eigenstate $Z'$ has
photon-like couplings to quarks, with $|f_n/f_p|\ll 1$.  On the other
hand, for $\epsilon = 0$, \emph{i.e.}\ considering only mass mixing, the
resulting $Z'$ has $Z$-like coupling to quarks which leads to
$|f_n/f_p| \gg 1$.  If both parameters are non-zero, we can achieve
$f_n/f_p\sim -0.7$ as required to evade the XENON constraints.  In
Fig.~\ref{fig2} we plot the contour levels of $f_n/f_p$ with
$m_{Z'}=4$ GeV. The figure shows that the required ratio of proton to
neutron coupling can be achieved by adjusting the two parameters
appropriately.

Interestingly the required value of $f_n /f_p \simeq -0.7$ is achieved
for $\epsilon \simeq \delta m^2/m_{Z}^2 $, as can be seen in
Fig.~\ref{fig2}.  In this limit, provided $\epsilon \ll 1$, we have
$\xi \approx \epsilon (1 + \hat s_W)$ leading to
\begin{eqnarray}
  f_p \approx
  -{\hat{g} f_\chi^V \over 4 \hat c_{W} } {c_\xi^2\over c_\epsilon} t_\xi
  {1 \over m_{Z'}^2} \frac{3 \hat s_{W}}{(1 +\hat s_{W})} 
  \ ,      \quad
  f_n \approx
  {\hat{g} f_\chi^V \over 4 \hat c_{W} } {c_\xi^2\over c_\epsilon} t_\xi
  {1\over m_{Z'}^2}\left(1 - \frac{\hat s_{W}}{1 + \hat s_{W}}\right) \ .
\end{eqnarray}
The corresponding ratio of the couplings is $f_n/f_p\approx -1/3\hat
s_{W} \approx -0.7 $.  The value of $f_n /f_p$ is linearly sensitive
to a rescaling $\epsilon \to \alpha \, \delta m^2/m_{Z}^2$ with
$\alpha$ of ${\cal O}(1)$.  Of course the underlying DM model needs to
satisfy this relation in some natural way.

\subsection{Baryonic $Z'$}

Next we consider the case where the SM is charged under the new
$U(1)_X$ gauge group.  As discussed above, when we constrain the
leptons to be uncharged under $U(1)_X$, the unique possibility is a
baryonic $Z'$ with $U(1)_X\equiv U(1)_B$. Anomaly-free models that
have a light $Z'$ coupling directly to baryon number such that
$f_u^\mathrm{V}=f_d^\mathrm{V}$ have been considered \emph{e.g.} in
Refs.\cite{Carone:1994aa,Carone:1995pu}. In the framework of a
baryonic $Z'$ we can also have kinetic mixing, but there is no
$Z$--$Z'$ mixing induced by the SM Higgs at tree-level as it does not
carry baryon number \cite{Carone:1995pu}.  However, mass mixing can
still be induced at the non-renormalizable level via \emph{e.g.} the
dim-6 operator $h_1^\dagger D_\mu h_1 h'^\dagger D^\mu h'$
discussed above.

\begin{figure}[!htp]
\begin{center}
{
\includegraphics[width=.45\columnwidth]{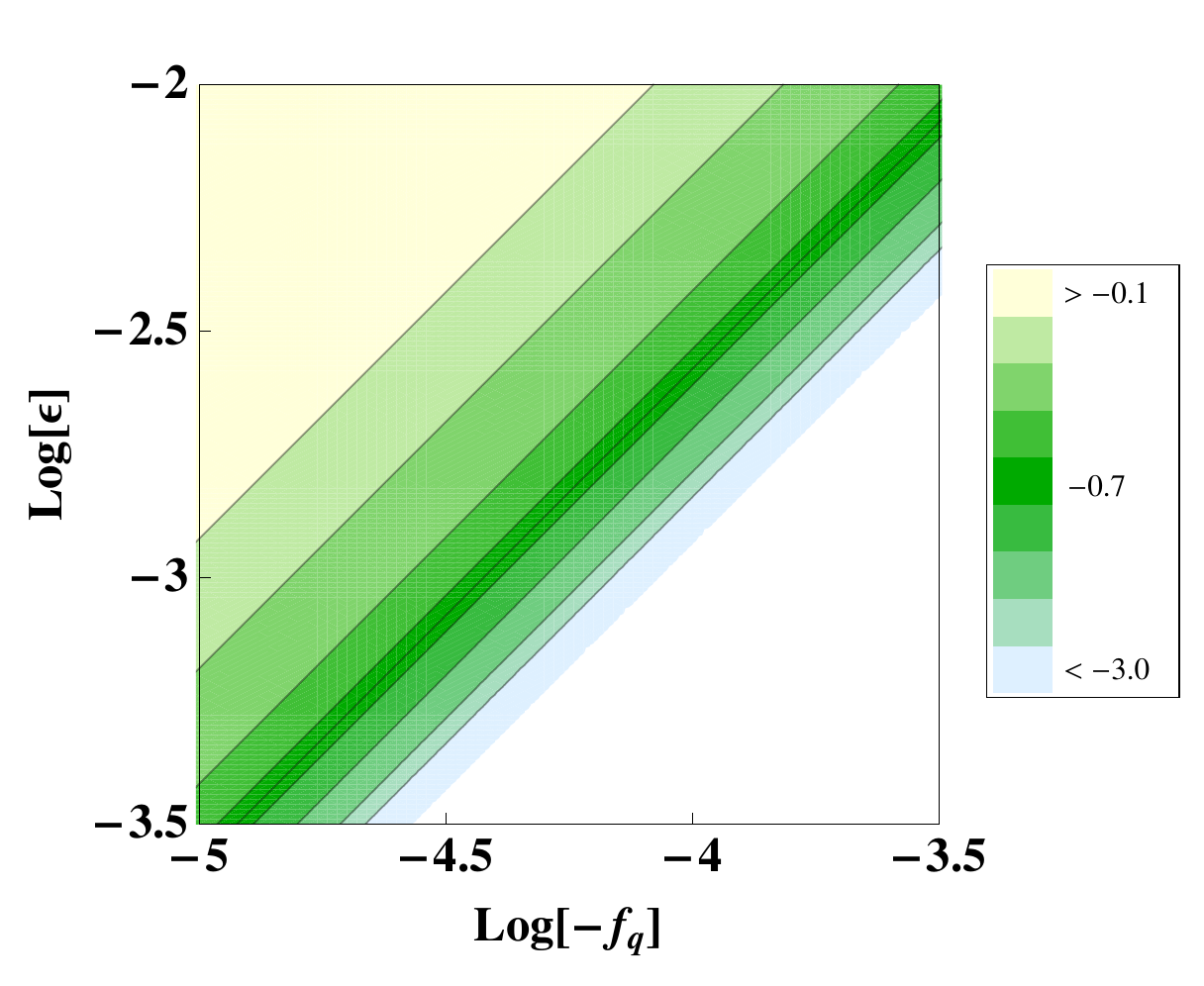}
\includegraphics[width=.45\columnwidth]{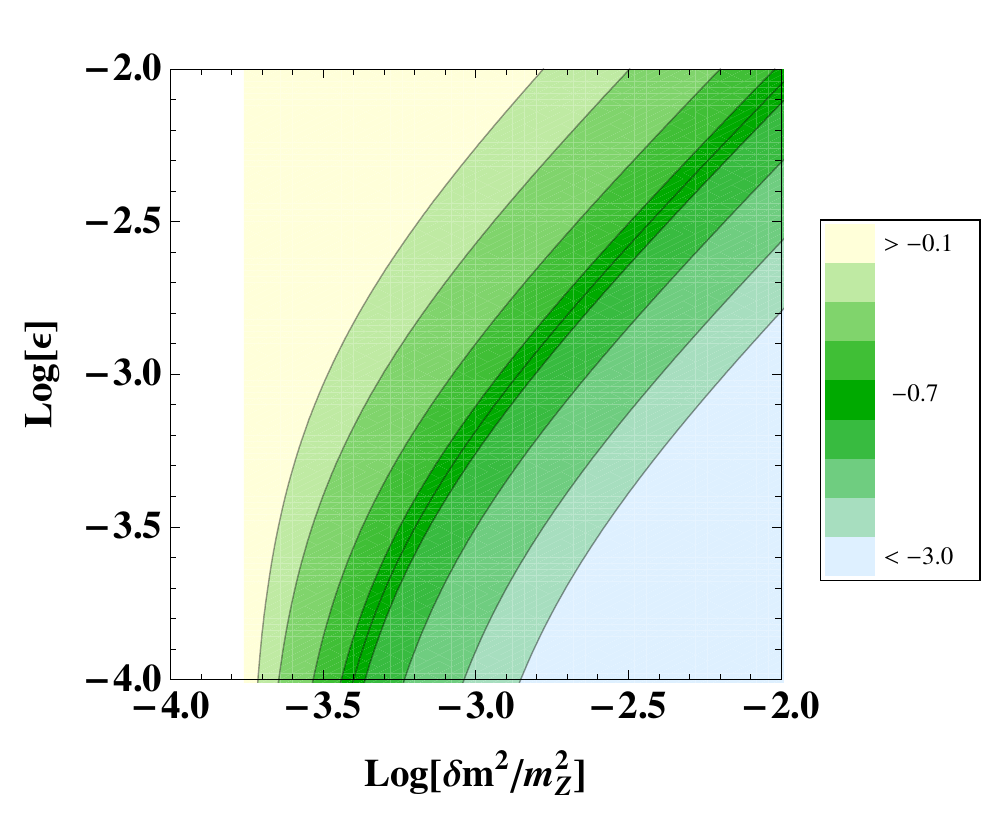}
}
\caption{Left: The ratio $f_n/f_p$ as a function of $f_q$ and
  $\epsilon$ for $\delta m^2 = 0$. Right: $f_n/f_p$ as a function of
  $\delta m^2$ and $\epsilon$ for $f_q^V = 10^{-5}$.}
\label{fig3}
\end{center}
\end{figure}

Since up and down quarks have equal charges under the baryonic $U(1)$,
large direct couplings will lead to $f_n/f_p \simeq 1$ for
$f_q^\mathrm{V}\gtrsim \epsilon$. In fact, $f_q^\mathrm{V}$ must be
over an order of magnitude smaller than $\epsilon$ to get $f_n/f_p
\simeq -0.7$, as shown in the left panel of Fig.~\ref{fig3}. As
discussed in the next section, the coupling $\epsilon$ is constrained
to be of order $10^{-2}$ or smaller, such that $f_q^\mathrm{V}\lesssim
10^{-3}$.

Since the direct quark couplings must be very small in order to obtain the desired ratio $f_n/f_p$, we can in fact relax the restriction that
the additional $U(1)$ must be baryonic. Allowing for couplings to leptons would facilitate the construction of an anomaly free model.

Charging the SM fermions under the new gauge group induces kinetic mixing via loops  \cite{Carone:1995pu}. Below the electroweak symmetry breaking scale the induced mixing of the $Z'$ with the $Z$ ($\epsilon_Z$) and the photon ($\epsilon_A$) will be different. At low energy one can then perform a transformation as in Eq.~(\ref{Zpmixing}) and recast the mixings $\epsilon_{A,Z}$ into a kinetic mixing $\epsilon$ and an additional $Z'$-$Z$ mass mixing term $\delta m^2$ as in Eq.~(\ref{LZprime}). The resulting mass mixing term will be of the same order as $\epsilon \, m_{Z}^2$, which was needed to achieve $f_n/f_p \approx -0.7$ in the absence of direct couplings to quarks. However, both these terms will be small compared to the direct coupling of $Z'$ to quarks, $f_{q}^V$, so that one actually obtains $f_p \approx f_n$ in the absence of additional contributions.
One possible way of avoiding the direct coupling contribution to $f_{n,p}$ is to couple the $Z'$ only to the second and third generation, so that the direct couplings do not contribute to $f_n / f_p$, but do induce kinetic mixings via quark loops. In this case, the required difference between $\epsilon_Z$ and $\epsilon_A$ can be achieved by imposing the initial condition $\epsilon_Z=\epsilon_A=0$ at the TeV scale.

\section{Limits on the mixing parameters}

In this section, we discuss various constraints on a light $Z'$
coupled to DM.  We do not require the $Z'$ interactions with the SM to
yield the correct thermal relic density for $\chi$ as the DM relic
density may well be of asymmetric origin. In particular the mass range
required to fit DAMA and CoGeNT is natural in models of asymmetric DM
(ADM), where the observed cosmological DM energy density is realised
for $m_{\chi} \sim 1-10$ GeV
\cite{Gelmini:1986zz,Kaplan:1991ah}. Moreover the ADM scenario avoids
the otherwise significant constraints from annihilation of light DM,
such as from the Sun \cite{Kappl:2011kz,Chen:2011vd}.

One class of relevant constraints arise from measurements of the
$Z$-pole --- these constrain directly the magnitude of the mixing
between $Z$ and $Z'$, while measurements of low energy observables are
sensitive to both the $Z'$ couplings and mass. Collider constraints
from mono-jet \cite{Goodman:2010yf,Goodman:2010ku,Bai:2010hh} and
di-jet searches \cite{Alitti:1990kw} also constrain the couplings to
quarks; however if the mediator is light, these constraints are fairly
weak.

\subsection{Electroweak precision  tests}

The constraints from electroweak precision measurements are encoded
in the $S$ and $T$ parameters. From the effective Lagrangian
formulation of the interaction
between $Z$ and the SM fermions \cite{Holdom:1990xp,Babu:1997st} 
\begin{align}
  \mathcal{L}_Z=&-\frac{e}{2 s_\mathrm{W}
    c_\mathrm{W}}\left(1+\frac{\alpha T}{2}\right) \bar
  f \gamma^\mu \nonumber \\
  &\times \left[\left(T_3^f-2Q^f\left(s_\mathrm{W}^2 + \frac{\alpha S
          - 4 c_\mathrm{W}^2 s_\mathrm{W}^2 \alpha
          T}{4(c_\mathrm{W}^2-s_\mathrm{W}^2)}\right)\right)-T^f_3\gamma^5\right]
  f Z_\mu \; ,
\end{align}
we can determine $S$ and $T$ to quadratic order in $\xi$ and
$\epsilon$,\footnote{Note that due to the requirement $f_n/f_p
  \simeq -0.7$, $\epsilon$ must be of the same order of magnitude as
  $\xi$. We therefore have a -2 appearing in the expression for $T$ as
  opposed to a -1 in \cite{Babu:1997st}. }
\begin{align}
  \alpha S = & 4 c_\mathrm{W}^2 s_\mathrm{W} \xi (\epsilon - s_\mathrm{W} \xi) 
  \; , \\
  \alpha T = & \xi^2\left(\frac{m_{Z'}^2}{m_{Z}^2}-2\right)+2
  s_\mathrm{W} \xi \epsilon \; ,
\end{align}
where $\alpha=e^2/4\pi$.  In the framework that we consider, one
typically has $\epsilon > s_\mathrm{W} \xi$ so that the $S$ parameter
is slightly positive. On the other hand, we require $m_{Z'} < m_{Z}$
so that the $T$ parameter will generally be negative. In this
direction, $S$ and $T$ are tightly constrained; however, given that in
our preferred parameter region we have $\xi \sim\epsilon\sim 0.01$
(see Sec.~\ref{sec:directdetection}), we get $S \simeq 0.01$,
$T \simeq -0.015$ which is adequately within the current constraints
\cite{Nakamura:2010zzi}.

The bound on $\xi$ and $\epsilon$ also implies that the $\rho$
parameter, given by
\begin{align}
  \rho = \frac{m_\mathrm{W}^2}{m_Z^2 c_\mathrm{W}^2} =
  \frac{s_\mathrm{W}^2}{\hat s_\mathrm{W}^2} \; ,
\end{align}
is within experimental uncertainties. To quadratic order in $\xi$ and
$\epsilon$ we obtain
\begin{align}
  \rho -1 = \frac{c_\mathrm{W}^2 \xi^2}{c_\mathrm{W}^2 - s_\mathrm{W}^2} 
  \left(\frac{m_{Z'}^2}{m_{Z}^2}-1\right) \; ,
\end{align}
which gives $\rho - 1 = -3 \times 10^{-4}$ for $\xi = 0.015$.

\subsection{$Z$ decay width}

The induced coupling of the $Z$ to the DM particle increases the
invisible $Z$ decay width. The contribution is approximately
\begin{align}
  \Gamma(Z \rightarrow \bar{\chi} \chi) = \frac{G_\mathrm{F}
    m_{Z}^3}{24 \pi \sqrt{2}} \left(|g_{\chi Z}^\mathrm{V}|^2 +
    |g_{\chi Z}^A|^2\right) = \frac{G_\mathrm{F} m_{Z}^3}{24 \pi
    \sqrt{2}} s_\xi^2 \left(|f_{\chi}^\mathrm{V}|^2 + |f_{\chi}^A|^2\right)
  \; .
\end{align}
Consequently, as long as $s_\xi < 0.015$, we satisfy the experimental
limit $\Gamma(Z \rightarrow \bar{\chi} \chi) < 1.5 \mev$ even if we
saturate the perturbative bound.

\subsection{Muon $g-2$}

In the presence of a new vector mediator the anomalous magnetic moment
of the muon will generally change. In the case of a dark $Z'$ with
$m_{Z'} \ll m_{Z}$, the contribution to $a_\mu = (g_\mu - 2)/2$ can
be estimated to be \cite{Chun:2010ve}
\begin{align}
  \delta a_\mu \simeq \frac{\alpha \xi^2}{3 \pi c_\mathrm{W}^2
    s_\mathrm{W}^2}  \frac{m_\mu^2}{m_{Z'}^2}
\end{align}
up to a factor of order unity\footnote{Here we assume that there are
  no cancellations between the axial and the vectorial part}. The
requirement $\delta a_\mu \simeq 4 \times 10^{-9}$ then implies an
approximate limit of \mbox{$m_{Z'} > 1 \gev$} for $\xi \simeq
10^{-2}$.

\subsection{Atomic parity violation}

The contribution of the $Z'$ to atomic parity violation (APV) is
proportional to the product of the axial coupling of the $Z'$ to
electrons, $g^A_{e Z'}$, and the vector coupling to quarks
$g^\mathrm{V}_{q Z'}$. In practice, however, one can only measure the
vector coupling to an entire nucleus, which is proportional to $Z f_p
+ (A-Z) f_n$. Because $f_p$ and $f_n$ have different signs in all the
cases that we consider, the two contributions nearly cancel, so that
no relevant constraint arises from APV. Note that in fact, the
strongest bounds on APV come from measurements of cesium, which has
almost the same ratio of protons to neutrons as xenon
(see Fig.~\ref{fig1}). Consequently, if indeed $f_n/f_p \simeq
-0.7$, the $Z'$ will give almost no contribution to APV in cesium.

\subsection{Hadronic decays}

Measurements of the decays of $\psi$ and $\Upsilon$ strongly constrain
the axial coupling of the $Z'$ to $c$ and $b$ quarks.  According to
Ref.\cite{Fayet:2007ua}, these limits are
\begin{align}
|g_{c}^A| & \lesssim 1.5 \times 10^{-3} \frac{m_{Z'}}{\text{GeV}} \ ,
\\
|g_{b}^A| & \lesssim 0.8 \times 10^{-3} \frac{m_{Z'}}{\text{GeV}} \ .
\end{align}
In our model, there is no direct axial coupling of the $Z'$ to quarks,
so that such a coupling can only arise from $Z-Z'$ mixing.  The
resulting coupling constants are proportional to $\xi $, \emph{i.e.}\ small
enough, given the previous constraints on $\xi$, if $m_{Z'} > 1$
GeV. The constraint for vector couplings is significantly weaker
(especially, if the decay of the $\Upsilon$ into two dark matter
particles is not possible). In fact, according to
Ref.\cite{Aranda:1998fr} it is easily possible to have
$|g_q^\mathrm{V}| \sim 0.1$. Ongoing and forthcoming searches for a
light $Z'$ at collider experiments such as BaBar, Belle, BEPC and
LHCb will be able to constrain these parameters more
tightly~\cite{Reece:2009un}.

\section{Dark matter direct detection}
\label{sec:directdetection}

\subsection{Spin-independent interactions}

We have previously determined best-fit parameters for a DM explanation
of the observed DAMA and CoGeNT modulations with SI scattering,
assuming that the relation $f_n/f_p = -0.7$ holds
\cite{Frandsen:2011ts}. The results were $m_\chi \sim 10$ GeV and
$\sigma_\mathrm{SI}\sim 10^{-38}- 10^{-37} \ \rm{cm}^2$ (depending on
whether a small inelasticity is included). We will now demonstrate
that such a large cross-section \emph{can} be realized while evading
all the limits discussed above.

First, we consider the case of a dark $Z'$ with no direct couplings to
SM quarks. In this case, there are 4 free parameters in our model:
$m_{Z'}$, $\delta m$, $m_\chi$ and $f^\mathrm{V}_\chi$.\footnote{The
  kinetic mixing $\epsilon$ is fixed by the requirement $f_n/f_p
  \simeq -0.7$, while $\xi$ is fixed as soon as $\delta m$ and
  $m_{Z'}$ are chosen.} Note that in the mass range $5 < m_\chi <
15$~GeV, the DM-neutron cross-section depends only very weakly on the
DM mass. As an example we take $m_{Z'} = 4$~GeV, $\delta m = 8$~GeV
and $m_\chi = 8$~GeV, leading to $\epsilon = 0.007$ and $\xi =
0.011$. The resulting cross-section is
\begin{align}
\sigma_n \simeq 8 \cdot 10^{-37} (f^\mathrm{V}_\chi)^2 \text{ cm}^2 \; .
\end{align}
Thus, a sufficiently large cross section can be achieved for $f^\mathrm{V}_\chi \sim 0.1$. Note that
there is no reason why $f^\mathrm{V}_\chi$ cannot be as large as its
perturbative bound $f^\mathrm{V}_\chi \lesssim \sqrt{4\pi}$. In fact,
if the DM sector is strongly interacting, $f^\mathrm{V}_\chi$ could be
even larger.  
By making $f^\mathrm{V}_\chi$ larger, one could still obtain cross sections of the correct magnitude even if the mixing parameters were smaller or $m_{Z'}$ larger than assumed above.
Note however, that since $\sigma_n \propto m_{Z'}^{-4}$, the
cross-section quickly becomes too small if $m_{Z'}$ is heavier than
$\sim 15 \gev$.

\subsubsection*{Baryonic $Z'$}

We now turn to the case of a baryonic $Z'$. As we have seen before,
direct couplings greater than $10^{-3}$ generally spoil the ratio
$f_n/f_p$. Thus, we must assume that the $Z'$ is extremely weakly
coupled, meaning that for charges of $\mathcal{O}(1)$ the gauge
coupling $g_X \ll 1$.  Setting $\delta m^2 = 0$ and $f^\mathrm{V}_q =
f^\mathrm{V}_\chi = 10^{-3}$ yields $f_n/f_p = -0.7$ if $\epsilon =
0.0265$ and $\xi = 0.0128$, which is still consistent with all
constraints. However, the resulting cross-section is 
too small: $\sigma_n < 10^{-41} \, \text{cm}^2$. Allowing a
non-zero $\delta m^2$ does not improve the situation significantly.
Consequently, if we wish to achieve a sufficiently high cross-section,
the coupling of the $Z'$ to the DM particle has to be significantly
larger than to quarks, $f_\chi^V \gg f_q^V$. While this might seem
unnatural in a `standard' $Z'$ model, it may well be possible in a
framework that derives from new strong dynamics, such as in
Ref.~\cite{Frandsen:2011kt}. We are investigating this possibility.

\subsection{Spin-dependent interactions}

If the $Z'$ couples to the axial DM current, there is an effective
axial-axial coupling between the DM particle and quarks given by
\begin{align}
b^\mathrm{A}_{q} = \frac{g^\mathrm{A}_{\chi}  g^\mathrm{A}_{q}}{m_{Z'}^2} \; ,
\end{align}
neglecting the contribution of the $Z$, which gives a correction of
order 1\%.

The quark-level couplings $b^\mathrm{A}_{q}$ induce the effective nucleon
couplings $a_{p,n}$ according to
\begin{align}
a_{p,n} = \sum_{q=u,d,s} \Delta q^{(p, n)} b^\mathrm{A}_{q} \; ,
\end{align}
where \cite{Ellis:2008hf}
\begin{align}
\Delta u^{(p)} & = \Delta d^{(n)} = 0.84 \pm 0.03 \nonumber \\
\Delta d^{(p)} & = \Delta u^{(n)} = -0.43 \pm 0.03 \nonumber \\
\Delta s^{(p)} & = \Delta s^{(n)} = -0.09 \pm 0.03 \; .
\end{align}
We assume that there is no direct axial coupling, \emph{i.e.}\
$f_u^\mathrm{A} = f_d^\mathrm{A} = 0$. In that case, according to
Eq.~\eqref{couplings}, $b_{d}^\mathrm{A} = b_{s}^\mathrm{A} = -
b_{u}^\mathrm{A}$. Consequently,
\begin{align}
a_{p} &= -1.36 b_{d}^\mathrm{A} , \nonumber \\
a_{n} &= 1.18 b_{d}^\mathrm{A}\; ,
\end{align}
meaning that the DM particle couples with roughly the same strength
but \emph{opposite} sign to protons and neutrons.

For zero momentum transfer, the spin-dependent (SD) DM-nucleon cross
section $\sigma^\mathrm{SD}_{p,n}$ is simply given by
\begin{equation}
\sigma^\mathrm{SD}_{p,n} = \frac{3}{\pi} \mu^2_{\chi (p,n)} a_{p,n}^2 \;.
\end{equation}
Using the results from Sec.~\ref{sec:zprime}, we get
\begin{align}
  \sigma^\mathrm{SD}_{p,n} & \simeq 0.1
  \frac{\mu^2_{\chi}}{m_{Z'}^4} {\hat g'^2} (f_\chi^\mathrm{A})^2 \left(
    \tfrac{c_\xi}{c_\epsilon}\right)^2 \left(\hat s_\mathrm{W} s_\xi
    - c_\xi t_\epsilon + \tfrac{\hat c_\mathrm{W}^2}{\hat s_\mathrm{W}} 
     s_\xi \right)^2  \nonumber \\
  & \simeq 3 \cdot 10^{-36}\text{cm}^2 (f_\chi^\mathrm{A})^2 \; ,
\end{align}
where in the last line we have substituted the same benchmark
parameters as above. We observe that the SD cross-section is
significantly larger than the SI one. 
It was shown~\cite{Savage:2004fn} (see also Ref.\cite{Schwetz:2011xm}) that a
SD DM-proton cross section of $\sim 10^{-36}$cm$^2$ is sufficient to
explain the DAMA annual modulation. Such a cross section can easily be obtained from a light $Z'$ mediator.

Consequently, a dark $Z'$ generically gives rise to both SI and SD
interactions of dark matter with nuclei. In the particular case that
we consider, SI interactions do not benefit from an enhancement proportional to $\mathcal{A}^2$, so that for 
 sufficiently large axial couplings both interactions should give similar
signals in direct detection experiments. This allows for the very
interesting possibility that the CoGeNT signal arises from SI
interactions, while the DAMA signal arises partly from SD
interactions.

\section{Conclusion}

The apparent conflict between the dark matter interpretation of the
DAMA and CoGeNT signals and the null results from other experiments is
a challenge for our understanding of dark matter. It has highlighted
the fact that various assumptions, which are made in analysing the
experimental data in order to derive the DM scattering cross-section
(or upper limits thereon), may be inappropriate. In particular, the DM
particle need not couple equally to protons and neutrons. Specifically
a ratio of neutron to proton coupling $f_n/f_p \simeq -0.7$ can reduce
the tension between experiments using different target materials.

In this paper, we have demonstrated that negative values of $f_n/f_p$
occur naturally if dark matter interactions with the experimental
targets are mediated by a GeV scale $Z'$ arising from a new $U(1)$
gauge group extension of the SM. There is a viable parameter region in
this model that leads to the desired value for $f_n/f_p$ and at the
same time gives sufficiently high cross sections to explain the
recently observed signals.

An interesting feature of light $Z'$ mediators is that they can also yield
a spin-\emph{dependent} cross-section which 
is sufficiently large to
account for the absolute signal
level observed by DAMA.
Forthcoming measurements of the $\Upsilon$ branching ratios will constrain the
mass and couplings of the $Z'$. Moreover, if the dark matter is
\emph{asymmetric}, such a large cross-section will affect heat
transport in the Sun (which has been sweeping up dark matter for
several billion years) and measurably alter the fluxes of low energy
Solar neutrinos
\cite{Frandsen:2010yj,Cumberbatch:2010hh,Taoso:2010tg}, thus providing
another diagnostic.

\section*{Acknowledgements}

We thank John March-Russell, Chris McCabe, Graham Ross and Stephen
West for useful discussions and the European Research and Training
Network ``Unification in the LHC era'' (PITN-GA-2009-237920) for
partial support. FK is supported by the DAAD and KSH acknowledges
support from ERC Advanced Grant BSMOXFORD 228169. This work was 
finalised during the CERN TH-Institute DMUH'11 and we thank all participants 
for stimulating discussions.

\bibliographystyle{ArXiv}
\bibliography{Zp}

\providecommand{\bysame}{\leavevmode\hbox to3em{\hrulefill}\thinspace}
\begin{thebibliography}{10}

\bibitem{Primack:1988zm}
J.~R. Primack, D.~Seckel, and B.~Sadoulet, Ann.Rev.Nucl.Part.Sci. \textbf{38}
  (1988), 751--807, Revised version.

\bibitem{Gaitskell:2004gd}
R.~Gaitskell, Ann.Rev.Nucl.Part.Sci. \textbf{54} (2004), 315--359.

\bibitem{Bernabei:2010mq}
R.~Bernabei, P.~Belli, F.~Cappella, R.~Cerulli, C.~Dai, et~al., Eur.Phys.J.
  \textbf{C67} (2010), 39--49,  [1002.1028].

\bibitem{Aalseth:2010vx}
CoGeNT, C.~E. Aalseth et~al.,  (2010),  1002.4703.

\bibitem{Aalseth:2011wp}
C.~Aalseth, P.~Barbeau, J.~Colaresi, J.~Collar, J.~Leon, et~al.,  (2011),
  1106.0650.

\bibitem{Kelso:2010sj}
C.~Kelso and D.~Hooper, JCAP \textbf{1102} (2011), 002,  [1011.3076].

\bibitem{Ahmed:2010wy}
CDMS-II Collaboration, Z.~Ahmed et~al., Phys.Rev.Lett. \textbf{106} (2011),
  131302,  [1011.2482].

\bibitem{Angle:2011th}
XENON10 Collaboration, J.~Angle et~al.,  (2011),  1104.3088.

\bibitem{Aprile:2011hi}
XENON100 Collaboration, E.~Aprile et~al., Phys.Rev.Lett. (2011),  [1104.2549].

\bibitem{Giuliani:2005my}
F.~Giuliani, Phys.Rev.Lett. \textbf{95} (2005), 101301,  [hep-ph/0504157].

\bibitem{Kurylov:2003ra}
A.~Kurylov and M.~Kamionkowski, Phys. Rev. \textbf{D69} (2004), 063503,
  [hep-ph/0307185].

\bibitem{Chang:2010yk}
S.~Chang, J.~Liu, A.~Pierce, N.~Weiner, and I.~Yavin, JCAP \textbf{1008}
  (2010), 018,  [1004.0697].

\bibitem{Feng:2011vu}
J.~L. Feng, J.~Kumar, D.~Marfatia, and D.~Sanford,  (2011),  1102.4331.

\bibitem{Frandsen:2011ts}
M.~T. Frandsen et~al.,  (2011),  1105.3734.

\bibitem{DelNobile:2011je}
E.~Del~Nobile, C.~Kouvaris, and F.~Sannino,  (2011),  1105.5431.

\bibitem{Schwetz:2011xm}
T.~Schwetz and J.~Zupan,  (2011),  1106.6241.

\bibitem{Farina:2011pw}
M.~Farina, D.~Pappadopulo, A.~Strumia, and T.~Volansky,  (2011),  1107.0715, *
  Temporary entry *.

\bibitem{Fox:2011px}
P.~J. Fox, J.~Kopp, M.~Lisanti, and N.~Weiner,  (2011),  1107.0717, * Temporary
  entry *.

\bibitem{McCabe:2011sr}
C.~McCabe,  (2011),  1107.0741.

\bibitem{Reece:2009un}
M.~Reece and L.-T. Wang, JHEP \textbf{0907} (2009), 051,  [0904.1743].

\bibitem{Belyaev:2010kp}
A.~Belyaev, M.~T. Frandsen, S.~Sarkar, and F.~Sannino, Phys.Rev. \textbf{D83}
  (2011), 015007,  [1007.4839].

\bibitem{Holdom:1985ag}
B.~Holdom, Phys.Lett. \textbf{B166} (1986), 196.

\bibitem{Babu:1997st}
K.~Babu, C.~F. Kolda, and J.~March-Russell, Phys.Rev. \textbf{D57} (1998),
  6788--6792,  [hep-ph/9710441].

\bibitem{Cassel:2009pu}
S.~Cassel, D.~M. Ghilencea, and G.~G. Ross, Nucl. Phys. \textbf{B827} (2010),
  256--280,  [0903.1118].

\bibitem{Hook:2010tw}
A.~Hook, E.~Izaguirre, and J.~G. Wacker,  (2010),  1006.0973.

\bibitem{Mambrini:2010dq}
Y.~Mambrini, JCAP \textbf{1009} (2010), 022,  [1006.3318].

\bibitem{Kang:2010mh}
Z.~Kang, T.~Li, T.~Liu, C.~Tong, and J.~M. Yang, JCAP \textbf{1101} (2011),
  028,  [1008.5243].

\bibitem{Chun:2010ve}
E.~J. Chun, J.-C. Park, and S.~Scopel, JHEP \textbf{1102} (2011), 100,
  [1011.3300].

\bibitem{Fox:2011qd}
P.~J. Fox, J.~Liu, D.~Tucker-Smith, and N.~Weiner,  (2011),  1104.4127.

\bibitem{Gondolo:2011eq}
P.~Gondolo, P.~Ko, and Y.~Omura,  (2011),  1106.0885.

\bibitem{Mambrini:2011dw}
Y.~Mambrini, JCAP \textbf{1107} (2011), 009,  [1104.4799].

\bibitem{Mambrini:2011pw}
Y.~Mambrini and B.~Zaldivar,  (2011),  1106.4819.

\bibitem{Feldman:2007wj}
D.~Feldman, Z.~Liu, and P.~Nath, Phys.Rev. \textbf{D75} (2007), 115001,
  [hep-ph/0702123].

\bibitem{Babu:1996vt}
K.~Babu, C.~F. Kolda, and J.~March-Russell, Phys.Rev. \textbf{D54} (1996),
  4635--4647,  [hep-ph/9603212].

\bibitem{Carone:1994aa}
C.~D. Carone and H.~Murayama, Phys.Rev.Lett. \textbf{74} (1995), 3122--3125,
  [hep-ph/9411256].

\bibitem{Carone:1995pu}
C.~D. Carone and H.~Murayama, Phys.Rev. \textbf{D52} (1995), 484--493,
  [hep-ph/9501220].

\bibitem{Gelmini:1986zz}
G.~Gelmini, L.~J. Hall, and M.~Lin, Nucl.Phys. \textbf{B281} (1987), 726.

\bibitem{Kaplan:1991ah}
D.~B. Kaplan, Phys. Rev. Lett. \textbf{68} (1992), 741--743.

\bibitem{Kappl:2011kz}
R.~Kappl and M.~W. Winkler, NUPHA,B850,505-521.2011 \textbf{B850} (2011),
  505--521,  [1104.0679].

\bibitem{Chen:2011vd}
S.-L. Chen and Y.~Zhang,  (2011),  1106.4044.

\bibitem{Goodman:2010yf}
J.~Goodman, M.~Ibe, A.~Rajaraman, W.~Shepherd, T.~M. Tait, et~al., Phys.Lett.
  \textbf{B695} (2011), 185--188,  [1005.1286].

\bibitem{Goodman:2010ku}
J.~Goodman, M.~Ibe, A.~Rajaraman, W.~Shepherd, T.~M. Tait, et~al., Phys.Rev.
  \textbf{D82} (2010), 116010,  [1008.1783].

\bibitem{Bai:2010hh}
Y.~Bai, P.~J. Fox, and R.~Harnik, JHEP \textbf{1012} (2010), 048,  [1005.3797].

\bibitem{Alitti:1990kw}
UA2, J.~Alitti et~al., Z. Phys. \textbf{C49} (1991), 17--28.

\bibitem{Holdom:1990xp}
B.~Holdom, Phys.Lett. \textbf{B259} (1991), 329--334.

\bibitem{Nakamura:2010zzi}
Particle Data Group, K.~Nakamura et~al., J.Phys.G \textbf{G37} (2010), 075021.

\bibitem{Fayet:2007ua}
P.~Fayet, Phys.Rev. \textbf{D75} (2007), 115017,  [hep-ph/0702176].

\bibitem{Aranda:1998fr}
A.~Aranda and C.~D. Carone, Phys.Lett. \textbf{B443} (1998), 352--358,
  [hep-ph/9809522].

\bibitem{Frandsen:2011kt}
M.~T. Frandsen, S.~Sarkar, and K.~Schmidt-Hoberg,  (2011),  1103.4350.

\bibitem{Ellis:2008hf}
J.~R. Ellis, K.~A. Olive, and C.~Savage, Phys. Rev. \textbf{D77} (2008),
  065026,  [0801.3656].

\bibitem{Savage:2004fn}
C.~Savage, P.~Gondolo, and K.~Freese, Phys.Rev. \textbf{D70} (2004), 123513,
  [astro-ph/0408346].

\bibitem{Frandsen:2010yj}
M.~T. Frandsen and S.~Sarkar, Phys. Rev. Lett. \textbf{105} (2010), 011301,
  [1003.4505].

\bibitem{Cumberbatch:2010hh}
D.~T. Cumberbatch, J.~A. Guzik, J.~Silk, L.~S. Watson, and S.~M. West, Phys.
  Rev. \textbf{D82} (2010), 103503,  [1005.5102].

\bibitem{Taoso:2010tg}
M.~Taoso, F.~Iocco, G.~Meynet, G.~Bertone, and P.~Eggenberger, Phys. Rev.
  \textbf{D82} (2010), 083509,  [1005.5711].

\end{thebibliography}

\end{document}